\documentclass[fleqn,twoside]{article}
\usepackage{espcrc2}
\usepackage[latin1]{inputenc}
\usepackage{graphicx}
\usepackage[figuresright]{rotating}

\newcommand{\AmS}{{\protect\the\textfont2
  A\kern-.1667em\lower.5ex\hbox{M}\kern-.125emS}}

\title{The AMS-02
Anticoincidence Counter}

\author{Ph. von Doetinchem\address[rwth]{I. Physics Institute B, RWTH Aachen University, Sommerfeldstr. 14, 52074 Aachen, Germany}, W. Karpinski$^a$,
Th. Kirn$^a$, K. Lübelsmeyer$^a$, St. Schael$^a$, M. Wlochal$^a$}
       
\begin{document}

\begin{abstract}
The AMS-02 detector will measure cosmic rays on the International Space Station. This contribution will cover production, testing, space qualification and integration of the AMS-02 anticoincidence counter. The anticoincidence counter is needed to to assure a clean track reconstruction for the charge determination and to reduce the trigger rate during periods of high flux.
\vspace{1pc}
\end{abstract}

\maketitle

\section{THE AMS-02 DETECTOR}

\begin{figure}[t]
\includegraphics[width=1.0\linewidth]{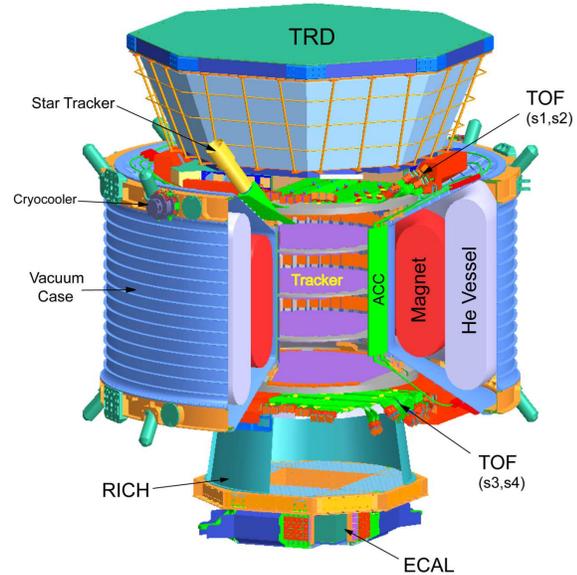}
\vspace{-1.3cm}
\caption{The AMS-02 detector.}
\label{f-ams_detector}
\end{figure}

The AMS-02 experiment will be installed on the International Space Station at an altitude of about 400\,km for about three years to measure cosmic rays without the influence of the Earth's atmosphere \cite{ams04}. The detector consists of several subdetectors for the determination of the particle properties, namely a transition radiation detector (TRD), a time of flight system (TOF), a cylindrical silicon microstrip tracker with eight layers surrounded by an anticoincidence counter system (ACC) in a superconducting magnet with a field of about 1\,T strength, a ring image \v{C}erenkov detector (RICH) and an electromagnetic calorimeter (ECAL) (fig.~\ref{f-ams_detector}).

\section{VETO SYSTEM -- THE ANTICOINCIDENCE COUNTER}

\begin{figure*}
\includegraphics[width=1.0\linewidth]{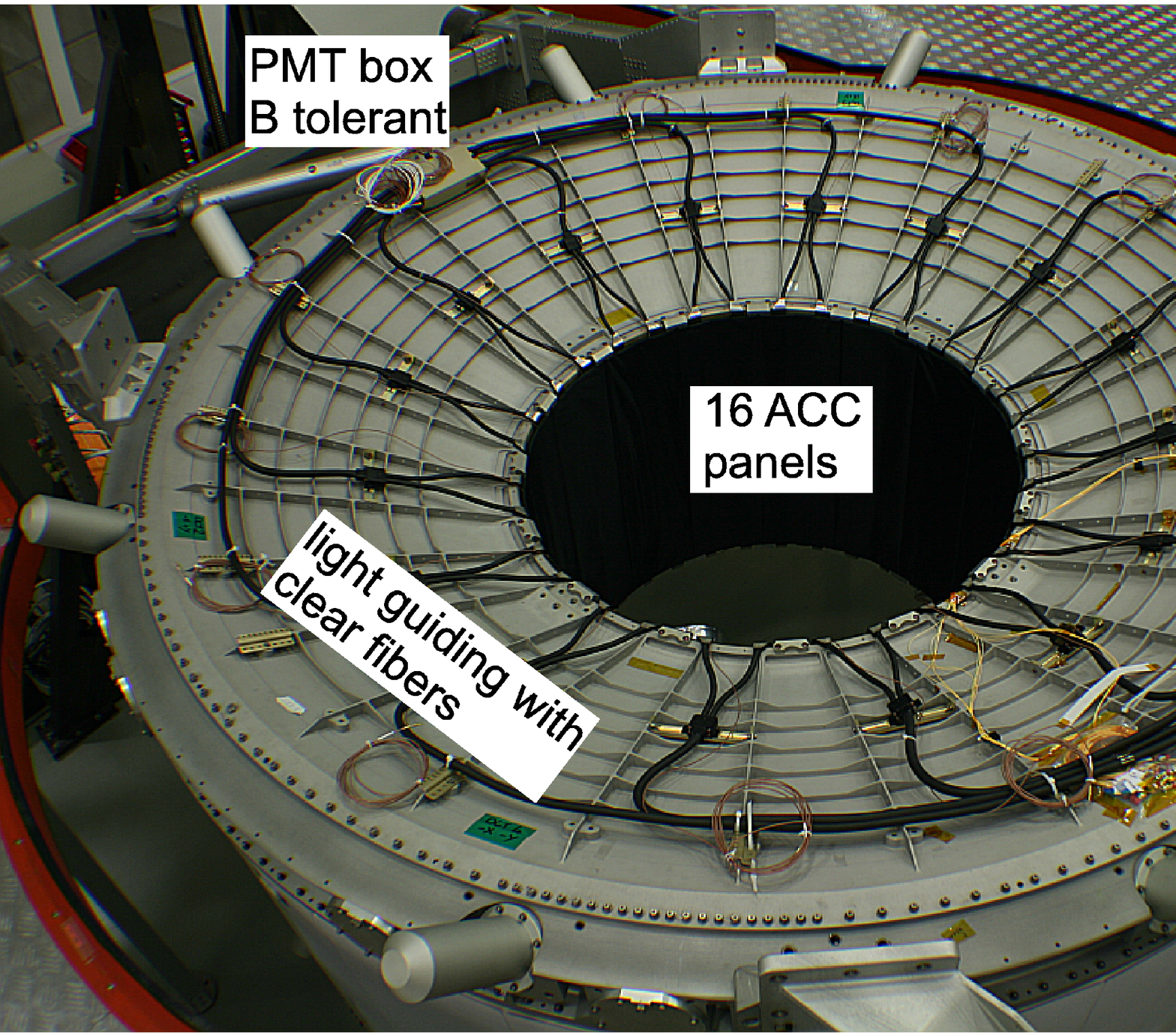}

\vspace{0.2cm}

\centerline{\includegraphics[width=0.7\linewidth]{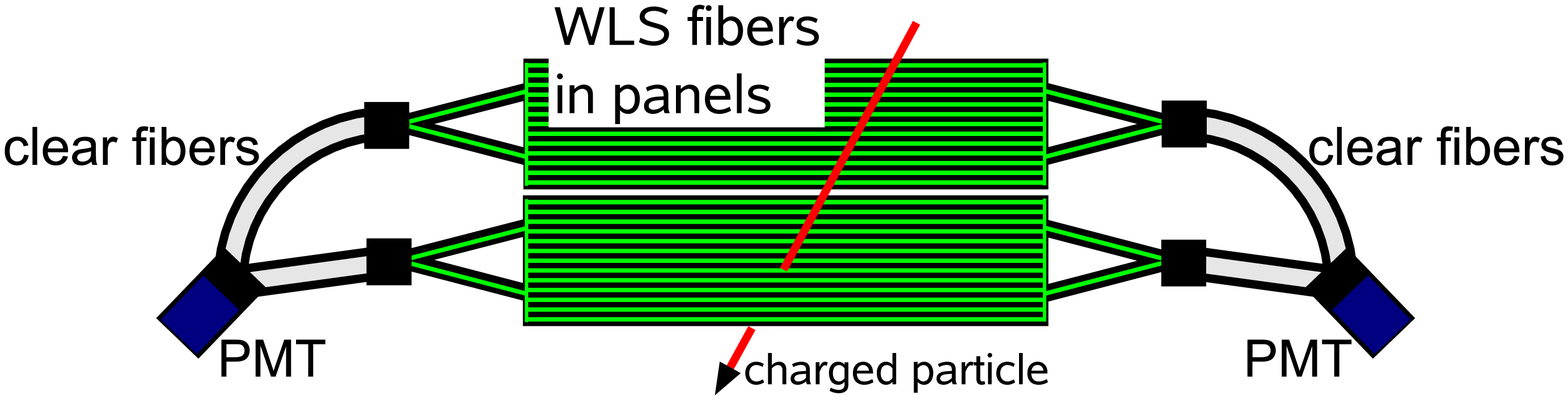}}
\vspace{-1.0cm}
\caption{\textbf{\textit{Upper)}} The anticoincidence counter system after integration (left) and the principle of component arrangement (right). \textbf{\textit{Lower)}} ACC working principle.}
\label{f-acc_overview}
\end{figure*}

The AMS-02 anticoincidence counter \cite{doe08} surrounds the silicon tracker with the purpose of vetoing to assure a clean track reconstruction (fig.~\ref{f-acc_overview}, upper). Particles entering the detector from the side or from secondary interactions inside it could distort the charge measurement. According to MC simulations to improve existing upper limits on antihelium an inefficiency of the ACC smaller than $10^{-4}$ is needed. The inefficiency is the ratio of missed to the total number of particle tracks crossing the ACC.

The second important task of the ACC is to reduce the trigger rate during periods of very large flux, e.g. in the South Atlantic Anomaly. The ACC detector will be used as a veto for the trigger decision. For that purpose, it is important to use a detector with a fast response. 

The ACC cylinder has a diameter of 1.1\,m and a height of 0.83\,m and is made out of 16 scintillation panels (Bicron BC-414) with a thickness of 8\,mm. The ultraviolet scintillation light ($\lambda\approx400$\,nm) through ionization losses of charged particles is absorbed by wavelength shifting fibers (WLS, Kuraray Y-11(200)M) which are embedded into the panels. The light is transformed to a wavelength of about 480\,nm and coupled to clear fiber cables (Toray PJU-FB1000) for the final transport to photomultiplier tubes (PMT, Hamamatsu R5946) with a quantum efficiency of about 20\,\% at 480\,nm \cite{ham94}. 

\subsection{Qualification of the Components}

The scintillator panels were tested after fabrication with reference PMTs without clear fiber coupling. The measured light yield for the 16 flight panels is on average 19 photo-electrons at the photocathode with an RMS of 1. 

Although the high magnetic field of the superconducting magnet is self-compensating and $B$-field tolerant fine mesh PMTs have been chosen, photomultiplier operation close to the panels would be too much distorted by the stray field. The signal has to be transported up to about 2\,m away from the scintillator to the PMT. The attenuation of the wavelength shifting fiber is large and a coupling to clear fibers can increase the total signal output. The requirements on the clear fiber are determined by the properties of the wavelength shifting fiber. The result of a near field measurement of the WLS fiber showed a homogeneous light output on the surface. The angular distribution determined in a far field measurement is quite wide because of non-ideal reflections and photon reabsorption in the fiber followed by isotropic fluorescence radiation at all angles. These effects are also responsible for the larger attenuation of the WLS fibers compared to the clear fibers. Therefore, it is important to match the angular acceptances of the fibers \cite{blu98}. The Toray PJU-FB1000 clear fiber has been chosen because it has a small attenuation at the green light of the WLS fiber ($\approx 0.1$\,dB/m) \cite{tor05} and a large angular acceptance. The average total damping of the clear fiber coupling and transportation is 2.1\,dB with an RMS of 0.1\,dB. 

The ACC is instrumented with 16 Hamamatsu R5946 photomultiplier tubes which are also used for the time of flight system. The fine mesh dynodes of these PMTs allow operation in magnetic fields without a large distortion of the electron trajectories \cite{ham94}. To even minimize this effect the PMTs are mounted parallel to the stray magnetic field of about 0.12\,T. The PMTs have undergone space qualification tests with temperature cycles in the non-operational range of -35°C to 50°C and the operational range of -30°C to 45°C and vibration tests with random frequencies and an average acceleration of 3.4\,$g$. The PMTs were tested with a reference panel and the selection criteria were the gain and the number of photo-electrons. Within about 5\,\% the PMTs do not show a variation before and after the space qualification tests.

\begin{figure}[t]
\includegraphics[width=1.0\linewidth]{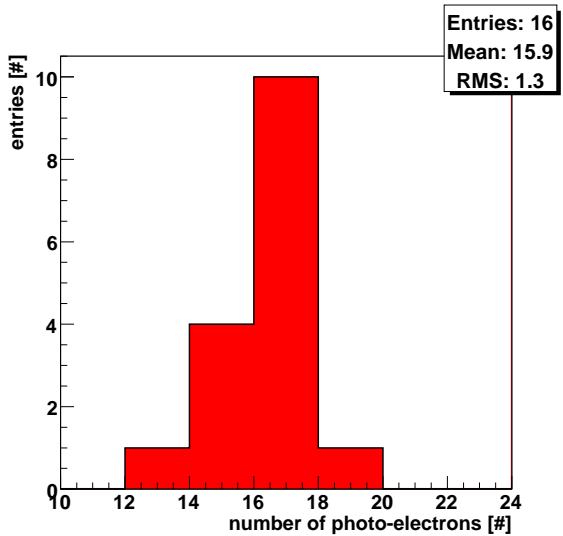}
\vspace{-1.3cm}
\caption{\label{f-acc_systemtest}Most probable number of photo-electrons for the 16 complete ACC panels}
\end{figure}

A set of two panels is read out by the same two photomultipliers one on top and one on the bottom, via clear fiber cables (Y-shape) in order to have redundancy and to save weight (fig.~\ref{f-acc_overview}, lower). The complete ACC system with the flight combination of panels, clear fiber cables and photomultipliers has a total weight of 53.7\,kg and an average output of ($16\pm1$) photo-electrons (fig.~\ref{f-acc_systemtest}). It was installed into the complete AMS-02 detector with a reproducibility for the mean signal output of $(99\pm8)$\,\%. 

The ACC signal is treated in two branches. The discriminator branch stores a time mark for every transaction above a threshold and can also be used for the quick veto decision of the level\,1 trigger generation. The ADC branch samples the pulse and measures the charge for each event. The discriminator threshold resolution is fine enough to resolve charges down to 0.7\,pC which corresponds to less than a photo-electron. This is important to reduce the trigger rate during periods of high flux.

\subsection{Overall Performance}

\begin{figure}[t]
\includegraphics[width=1.0\linewidth]{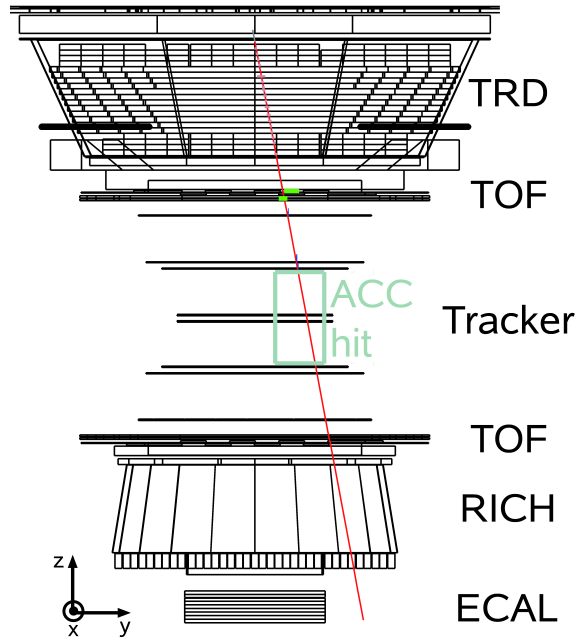}
\vspace{-1.0cm}
\caption{Clean event for the ACC analysis during the AMS-02 preintegration runs with atmospheric muons. The outline of the ACC hit shows only a part of a panel.}
\label{f-eventdisplay_1209556604_33077}
\end{figure}

\begin{figure}[t]
\includegraphics[width=1.0\linewidth]{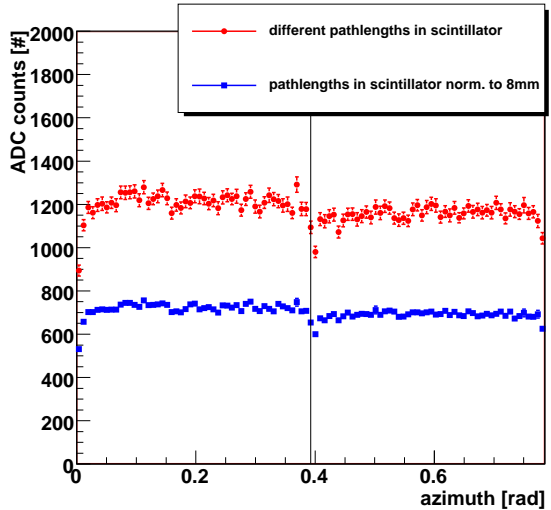}
\vspace{-1.3cm}
\caption{Mean ADC values vs. the azimuthal position in an ACC sector. The line indicates the slot between the two panels of the sector.}
\label{f-081125_4_10_3_0_1_2_3_acc_phi_mod_adc_highest_bothmean}
\end{figure}  

\begin{figure}[t]
\includegraphics[width=1.0\linewidth]{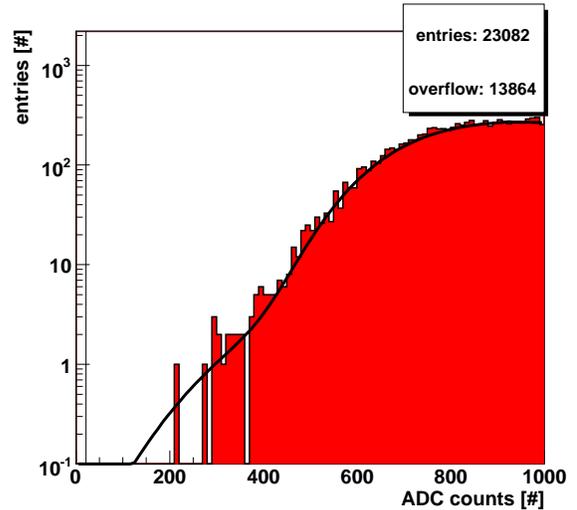}
\vspace{-1.3cm}
\caption{ADC values from atmospheric muons with signal model. The vertical line indicates the cut for the definition of a good ACC event.}
\label{f-081125_4_10_3_0_1_2_3_acc_adc_highest_zoom}
\end{figure} 

The analysis of the atmospheric muons collected with the whole AMS-02 experiment allowed the extraction of further properties. The inefficiency was calculated by using tracks in the TRD and the tracker (fig.~\ref{f-eventdisplay_1209556604_33077}). The tracks in the transition radiation detector were extrapolated to the silicon tracker. Hits in the tracker close to this track were used for a new fit with higher resolution. Tracks pointing to an ACC panel were analysed. Good ACC events show at least one ADC value above three RMS of the corresponding pedestal distribution. The slot regions between two scintillator panels which are realized with tongue and groove dominate the determination of the mean inefficiency. The average ACC PMT signal behavior as a function of the azimuthal angle is shown in fig.~\ref{f-081125_4_10_3_0_1_2_3_acc_phi_mod_adc_highest_bothmean}. The drop ($\approx20$\,\%) for the slot region between the panels of one sector sharing their PMTs is not as strong as for the slots between sectors ($\approx30$\,\%). The analysis did not show a single missed event (fig.~\ref{f-081125_4_10_3_0_1_2_3_acc_adc_highest_zoom}). This information can be used to set an upper limit on the ACC inefficiency of $I<1.3\cdot 10^{-4}\mbox{ @ 95\,\% confidence level}.$ In addition, a signal model derived from the different spectra in the slot and central regions is shown.

The result is statistically limited. From MC simulations we expect an even smaller inefficiency when the experiment is operated in space due to the different angular distribution of cosmic rays.

\section{CONCLUSION}

The production of a highly efficient veto counter was successful. The detector will be able to operate reliably in a Space environment and high magnetic field.
 
This project is funded by the German Space Agency DLR under contract No. 50OO0501.

\section{ACKNOWLEDGMENTS}

The authors wish to thank the following people for their
support with the fiber measurements A.~Bachmann and H.~Poisel and A.~Basili, V.~Bindi, V.~Choutko, E.~Choumilov, A.~Kounine and  A.~Lebedev for the help during data taking at CERN and with the AMS software.

\end{document}